\documentclass[12pt, a4paper]{article}
\usepackage[margin=1in]{geometry}
\usepackage{amsfonts, amsmath, amssymb, bm}
\usepackage{tabu}
\usepackage[utf8x]{inputenc}
\usepackage{enumerate}
\usepackage{tgcursor}
\usepackage{xcolor}
\usepackage{graphicx}
\usepackage{placeins}
\usepackage{listings}
\usepackage{stackengine}

\usepackage{caption}
\usepackage{subcaption}
\usepackage{mathptmx}
\usepackage{setspace}
\usepackage{footnote}
\usepackage{comment}
\usepackage{footmisc}
\usepackage{hyperref}
\numberwithin{equation}{section}
\usepackage{eucal}
\usepackage{verbatim}
\usepackage{mathrsfs}
\usepackage{natbib}% Citation support 
\newcommand{\va}{\mathbf{a}}

\newcommand{\vB}{\mathbf{B}}
\newcommand{\vc}{\mathbf{c}}
\newcommand{\vC}{\mathbf{C}}

\newcommand{\vF}{\mathbf{F}}

\newcommand{\vI}{\mathbf{I}}

\newcommand{\vs}{\mathbf{s}}

\newcommand{\vV}{\mathbf{V}}
\newcommand{\vw}{\mathbf{w}}

\newcommand{\vX}{\mathbf{X}}

\newcommand{\vz}{\mathbf{z}}

\newcommand{\vbe}{\boldsymbol{\beta}}

\newcommand{\vep}{\boldsymbol{\epsilon}}

\newcommand{\vmu}{\boldsymbol{\mu}}

\newcommand{\vSig}{\boldsymbol{\Sigma}}

\newcommand{\vtheta}{\boldsymbol{\theta}}

\newcommand{\vrho}{\boldsymbol{\rho}}

\newcommand{\vpsi}{\boldsymbol{\psi}}

\newcommand{\Gau}{\text{Gau}}

\hypersetup{
    colorlinks=true,
    linkcolor=blue,
    urlcolor=blue,
    citecolor=black
}

\newcommand{\ssection}[1]{%
  \section[#1]{\centering\normalfont\scshape #1}}
\newcommand{\ssubsection}[1]{%
  \subsection[#1]{\raggedright\normalfont\itshape #1}}
  
\newcounter{savefootnote}
\newcounter{symfootnote}
\newcommand{\symfootnote}[1]{%
   \setcounter{savefootnote}{\value{footnote}}%
   \setcounter{footnote}{\value{symfootnote}}%
   \ifnum\value{footnote}>8\setcounter{footnote}{0}\fi%
   \let\oldthefootnote=\thefootnote%
   \renewcommand{\thefootnote}{\fnsymbol{footnote}}%
   \footnote{#1}%
   \let\thefootnote=\oldthefootnote%
   \setcounter{symfootnote}{\value{footnote}}%
   \setcounter{footnote}{\value{savefootnote}}%
}

\begin{document}

%%%%%%%%%%%%%%%%%%%%%%%%%%%%%%%%%%%%%%%%%%%%%%%%%%%%%%%%%%%%%%%%%%%%%%
%%%% Title
%%%%%%%%%%%%%%%%%%%%%%%%%%%%%%%%%%%%%%%%%%%%%%%%%%%%%%%%%%%%%%%%%%%%%%
\begin{center}
   %\Large\centering\normalfont\scshape{On the machinations of the annual and semi-annual temperature harmonics}
   \Large\centering\normalfont\scshape{On the spatial and temporal shift in the archetypal seasonal temperature cycle as driven by annual and semi-annual harmonics}
   % \Large\centering\normalfont\scshape{Quantifying the effect of spatial and temporal variation of annual and semi-annual harmonics on minimum and maximum temperature across North America} 
\end{center}

%%%%%%%%%%%%%%%%%%%%%%%%%%%%%%%%%%%%%%%%%%%%%%%%%%%%%%%%%%%%%%%%%%%%%%
%%%% Authors
%%%%%%%%%%%%%%%%%%%%%%%%%%%%%%%%%%%%%%%%%%%%%%%%%%%%%%%%%%%%%%%%%%%%%%
\begin{center}
    Joshua S. North\footnote[1]{Corresponding author: joshuanorth@mail.missouri.edu, 146 Middlebush Hall, Columbia, MO 65211 }, Erin M. Schliep, Christopher K. Wikle\
    
    Department of Statistics, University of Missouri
\end{center}

\begin{abstract}
    Statistical methods are required to evaluate and quantify the uncertainty in environmental processes, such as land and sea surface temperature, in a changing climate.  Typically, annual harmonics are used to characterize the variation in the seasonal temperature cycle. However, an often overlooked feature of the climate seasonal cycle is the semi-annual harmonic, which can account for a significant portion of the variance of the seasonal cycle and varies in amplitude and phase across space. Together, the spatial variation in the annual and semi-annual harmonics can play an important role in driving processes that are tied to seasonality (e.g., ecological and agricultural processes). We propose a multivariate spatio-temporal model to quantify the spatial and temporal change in minimum and maximum temperature seasonal cycles as a function of the annual and semi-annual harmonics. Our approach captures spatial dependence, temporal dynamics, and multivariate dependence of these harmonics through spatially and temporally-varying coefficients. We apply the model to minimum and maximum temperature over North American for the years 1979 to 2018. Formal model inference within the Bayesian paradigm enables the identification of regions experiencing significant changes in minimum and maximum temperature seasonal cycles due to the relative effects of changes in the two harmonics.
    
    \begin{center}
        \textit{Key Words:} Spatio-Temporal Statistics, Dynamic System Modeling, Predictive Process, North American Temperature Cycle, Spatial Synchrony
    \end{center}
    
\end{abstract}

%%%%%%%%%%%%%%%%%%%%%%%%%%%%%%%%%%%%%%%%%%%%%%%%%%%%%%%%%%%%%%%%%%%%%%
\ssection{Introduction}
%%%%%%%%%%%%%%%%%%%%%%%%%%%%%%%%%%%%%%%%%%%%%%%%%%%%%%%%%%%%%%%%%%%%%%

% Discuss the impact of spatial and temporal synchrony of climate on living organisms
In ecology, ``spatial synchrony'' is a concept that describes coincident in time variations in an ecological process across geographically separated populations \citep{Liebhold2004}. In many cases, this synchrony leads to symbiotic relationships that are tied to environmental seasonal cycles. For example, an important climate-driven issue facing forest health concerns native bark beetle infestations, with current beetle outbreaks among the most severe in recorded history \citep{Bentz2010}. Historically, exposure to very cold temperatures is necessary to control the beetle population, but increasing seasonal minimum temperatures in northern latitudes has disrupted this historical synchrony, limiting beetle mortality, and increasing tree mortality.  Other examples of spatial synchrony being disrupted by changes in environmental seasonal cycles include ocean primary productivity \citep{Defriez2016}, lake stratification \citep{Kraemer2015}, migration patterns \citep{Usui2017}, and flood hazards \citep{Arnell2016}, among others.  The broad extent of these impacts highlight the importance of understanding how spatial patterns in environmental seasonal cycles vary through time.

The seasonal cycle in atmospheric variables is a direct response to the variation in solar insolation due to the Earth's orbital path around the Sun.  Specifically, the atmospheric response to the overhead sun crossing the equator twice a year suggests a more complicated seasonal variation that includes both an annual and a semi-annual harmonic.  Harmonic analysis has been used by meteorologists and climatologists to characterize the connection between these harmonics and the observed seasonal cycle since the early 20th century (e.g., see \citet{Hsu1976, Hsu1976b} for a review of this early work).  Although the semi-annual harmonic typically contributes less variance to the seasonal cycle than the annual harmonic in the Northern mid-latitudes, its amplitude and phase vary considerably across space, and there are regions in which it can significantly affect the seasonal cycle (e.g., shifting the phase, strengthening the peak, flattening the minimum; see \citet{White1978}).  One of the first studies to discuss a specific dynamical mechanism behind the semi-annual cycle was \citet{VanLoon1967}, in which he showed that the semi-annual cycle in the high Southern latitudes was the result of differential heating due to north--south land/sea contrasts between Antarctica and the Southern ocean.  Unlike the north--south contrast exhibited in high latitudes, \citet{Wikle1996} showed evidence that the Northern hemisphere extratropical semi-annual cycle exhibits a strong east--west structure and is governed by the spatio-temporal asymmetries in the seasonal variation of the northern hemisphere stationary eddies (e.g., the wave structure in the atmospheric circulation).  Because this variation in stationary eddies is due to east--west differential heating from land/sea contrasts, the fundamental mechanism for both the extra-tropical and high-latitude semi-annual cycles is due to land--sea contrasts and the impact this has on the atmospheric circulation.

It is well-known that atmospheric circulation patterns are varying due to differing responses of land and sea to climate forcing (e.g., \citet{Sutton2007}). This suggests that the annual and semi-annual harmonics are also likely varying.  \citet{Stine2009} investigated the change in the annual harmonic component of surface temperature between the years 1900-1953 and 1954-2007 by looking at the lag (difference between temperature and local solar insolation phases) and gain (ratio of temperature and insolation amplitudes).  Based on simple t-test comparisons of the lag and gain for the different time periods, they showed that the annual temperature cycle has changed, but asymmetrically across space.  \citet{Dwyer2012} also showed that there has been heterogeneous variation in the annual amplitude and phase of the mean surface temperature cycle in response to greenhouse gases.  These analyses did not explicitly consider the semi-annual component of the seasonal cycle, multivariate seasonal variation of atmospheric variables, nor did they consider a formal model-based uncertainty quantification framework that could accommodate spatial and temporal variation of the harmonics.

% Show that statistical climate models have the ability to bridge determine the impact of the semi-annual cycle
Dynamic spatio-temporal models (DSTMs) are well established in the literature for modeling complex spatial processes that evolve over time (see \citet{Cressie2011} for a collection of references and methods). Statistical DSTMs are able to capture spatial and temporal dependence in the process across different scales, while retaining the ability to capture uncertainty in parameter estimation and prediction.  Surface temperatures over land generally can be decomposed into three components: a term to account for trend, a seasonal component, and a ``weather'' component.  We would expect the first two of these components to vary somewhat slowly across time, with near-by locations experiencing similar temperatures and temperature variation through time, whereas the weather component corresponds to a dynamic process observed at finer spatial and temporal scales \citep{Wikle1998}. The most natural way to accommodate slowly varying time variation in trend and seasonal parameters is via the dynamic linear model (DLM) paradigm (e.g., \citet{West2006}).  Such models are commonly extended to the dynamic evolution of parameters in spatio-temporal and multivariate settings by representing the parameters as spatial fields that vary in time according to a DLM (e.g., see the overviews in \citet{Gelfand2010,Cressie2011, Banerjee2015, Gelfand2017}).

The main modeling contribution of this work is the development of a joint statistical framework for time-varying minimum and maximum temperature cycles, which are specified through the annual and semi-annual harmonics, while accounting for spatial and temporal dependence. 
This model is motivated by the fact that responses to changes in heating are asymmetric in space; thus, we expect that the annual and semi-annual harmonics in temperature are varying in time differently across space, leading to time-varying differences in seasonal cycles.
By adopting a Bayesian framework for parameter estimation for the associated DSTM model, we are able to quantify the extent to which regions across North America are experiencing significant shifts in the minimum and maximum temperature cycles.  Significant asymmetric shifts in minimum and maximum temperature seasonal cycles may seriously affect biological processes that are synchronously linked to such cycles.

The remainder of this manuscript is structured as follows.  In Section \ref{sec:motivation} we describe the data used in the analysis and provide a brief exploratory analysis to motivate the work. Section \ref{sec:methods} details the joint model specification using the annual and semi-annual harmonics and methods for model inference. Section \ref{sec:results} presents the findings of the analysis and Section \ref{sec:discussion} provides a discussion and directions for future work.

%%%%%%%%%%%%%%%%%%%%%%%%%%%%%%%%%%%%%%%%%%%%%%%%%%%%%%%%%%%%%%%%%%%%%%
\ssection{Data and Preliminary Analysis}
\label{sec:motivation}
%%%%%%%%%%%%%%%%%%%%%%%%%%%%%%%%%%%%%%%%%%%%%%%%%%%%%%%%%%%%%%%%%%%%%%

%%Here is the data
For the analyses presented here, we consider air temperature (deg C) data at two meters above the surface obtained from the National Center for Environmental Prediction (NCEP) Reanalysis\footnote{\href{https://www.esrl.noaa.gov/psd/}{https://www.esrl.noaa.gov/psd/}}.  The data are available at three hour intervals for each day from January 1, 1979 to December 31, 2018, which we summarize as daily minimum and maximum temperature.
The data are on a $349 \times 277$ Northern Lambert Conformal Conic grid, with corners at approximately (1.000N, 145.500W), (0.898N, 68.320W), (46.354N, 2.570W), and (46.634N, 148.642E).
All data exploration was conducted on a reduced spatial domain with corners at approximately (16.103N, 140.543W), (15.997N, 73.229W), (57.601N, 22.274W), and (57.856N, 168.499E).

%%Here we say that we can use Fourier series to represent the temperature cycles. Then, that we only really want to look at the first 2
Let $z_t$ denote temperature (say, minimum or maximum) on day $t$ where $t= 1, \dots, T$, and $T$ is the number of days in the year (365 or 366 for leap years).
The Fourier series representation of the time series, expressed in amplitude-phase form, is given as
\begin{align}\label{eq:harm_amp_phi}
    z_t = a_0 + \sum_{h=1}^{T/2} A_h \cos \Big(\frac{2 \pi h t}{T} + \varphi_h \Big),
\end{align}
where $A_h$ and $\varphi_h$ are the amplitude and phase, respectively, for the $h^{th}$ harmonic component.
Reparameterizing Eqn. \ref{eq:harm_amp_phi} in terms of its Fourier coefficients results in
\begin{align}\label{eq:harm_exp}
    z_t = a_0 + \sum_{h=1}^{T/2} a_h \cos \Big( \frac{2 \pi t h}{T} \Big) + b_h \sin \Big( \frac{2 \pi t h}{T} \Big),
\end{align}
where $a_h$ and $b_h$ are the Fourier coefficients for the $h^{th}$ harmonic, related to the amplitude by $A_h = \sqrt{a^2_h + b^2_h}, \,\, A_h \in [0, \infty)$, and the phase by $\varphi_h = \tan^{-1}(-b_h/a_h), \, \varphi_h \in [-\pi/h, \pi/h]$.
We restrict our estimation to the first two harmonics ($h=1, 2$) as discussed in the introduction, and refer to a ``cycle'' as the sum of the first and second harmonics hereafter.

\begin{figure}[ht]
    \centering
    \includegraphics[width = \linewidth]{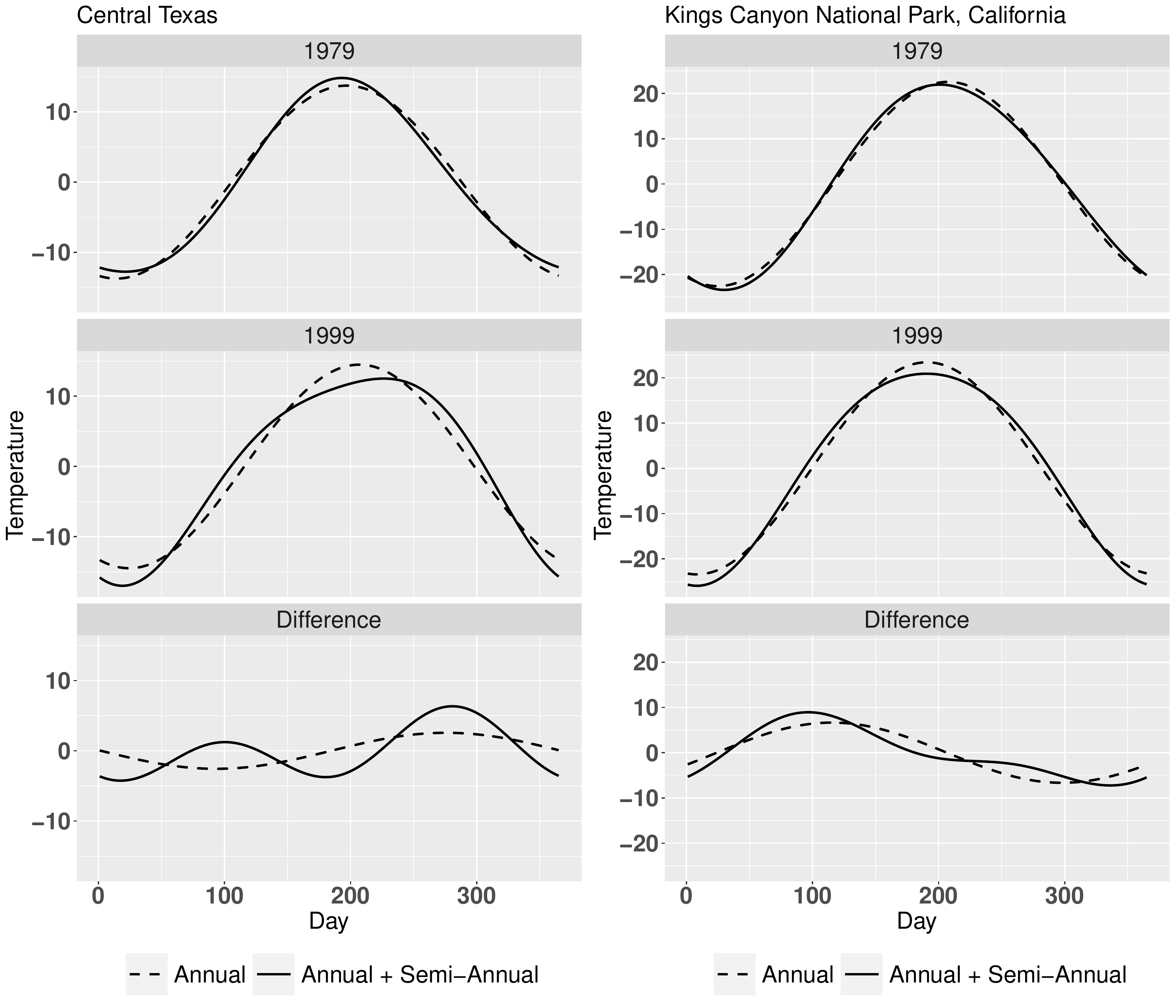}
    \caption{Comparison of estimated seasonal cycle for minimum temperature using the first and second Fourier harmonics (solid line) compared to just the first Fourier harmonic (dashed line) in central Texas and Kings Canyon National Park, California.  The top plot is the estimated cycle for 1979, the middle for 1999, and the bottom showing the difference (1999 year - 1979 year).}
    \label{fig:semi_just}
\end{figure}

%% Here we show how the cycles vary over time, and that the amount the semi-annual cycle effects the overall cycle is different for different years.
To investigate the possible relative importance of the semi-annual harmonic in North American minimum and maximum temperature cycles, Figure \ref{fig:semi_just} shows the estimated seasonal cycles for minimum temperature when both the annual and semi-annual harmonic components are included (solid) compared to those in which only the annual component is considered (dashed). These estimated cycles are shown for two different years, 1979 and 1999, and two different locations, one in central Texas and the other in Kings Canyon National Park, California. 
The top panel shows the estimated cycle for the year 1979, the middle panel for the year 1999, and the bottom panel shows the difference between the two cycles, with the estimates from the 1979 cycle subtracted from the 1999 cycle.
These plots illustrate the impact the semi-annual component can have on the temperature cycle, how the impact changes through time, and how these features vary across space.
The estimated cycles for 1979 are very similar for both locations, suggesting the semi-annual harmonic had little influence on the minimum temperature cycle at these locations.
Conversely, the temperature cycles for 1999 are much more dissonant for both locations, implying the semi-annual harmonic had a greater impact on the temperature cycle in 1999 than in 1979.
The impact of the semi-annual harmonic of minimum temperature  can be seen clearly in the bottom panel, where the difference between the estimated cycles between the two years using both annual and semi-annual components in central Texas shows a cyclical deviation from the difference in cycle estimates using only the annual component. At Kings Canyon,  the cycle with the annual and semi-annual components oscillates about the difference using only the annual component.
This suggests that the impact of the semi-annual component is spatially and temporally varying, and is important in capturing shifts in the temperature cycle through time.

\begin{figure}[ht]
    \centering
    \includegraphics[width = \linewidth]{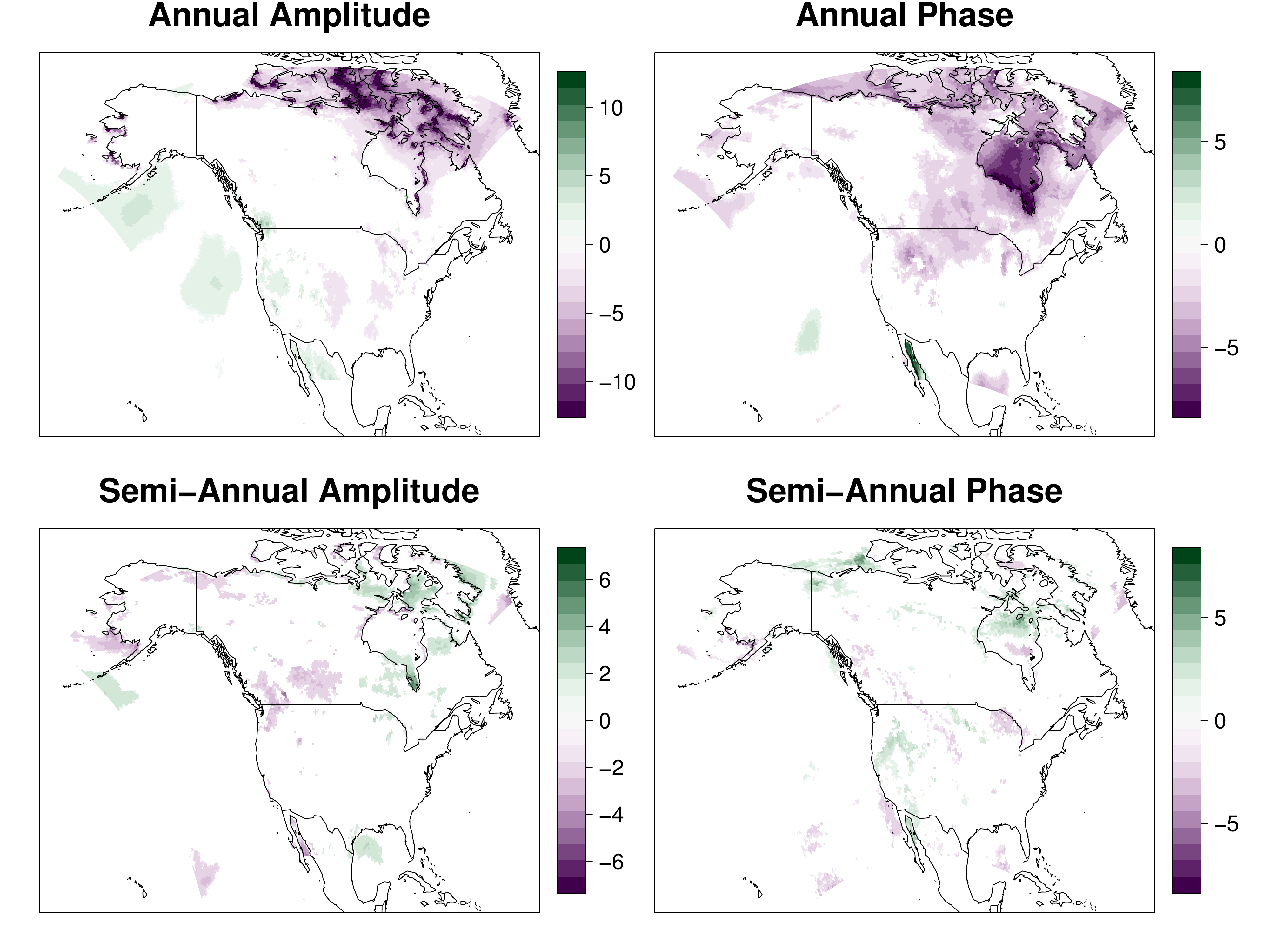}
    \caption{Computed t-statistics of the Fourier coefficients for the minimum temperature cycle. Locations with 95\% point-wise confidence intervals not including 0 are shown, with the color corresponding to the t-statistic value.}
    \label{fig:fftplots_min}
\end{figure}

\begin{figure}[ht]
    \centering
    \includegraphics[width = \linewidth]{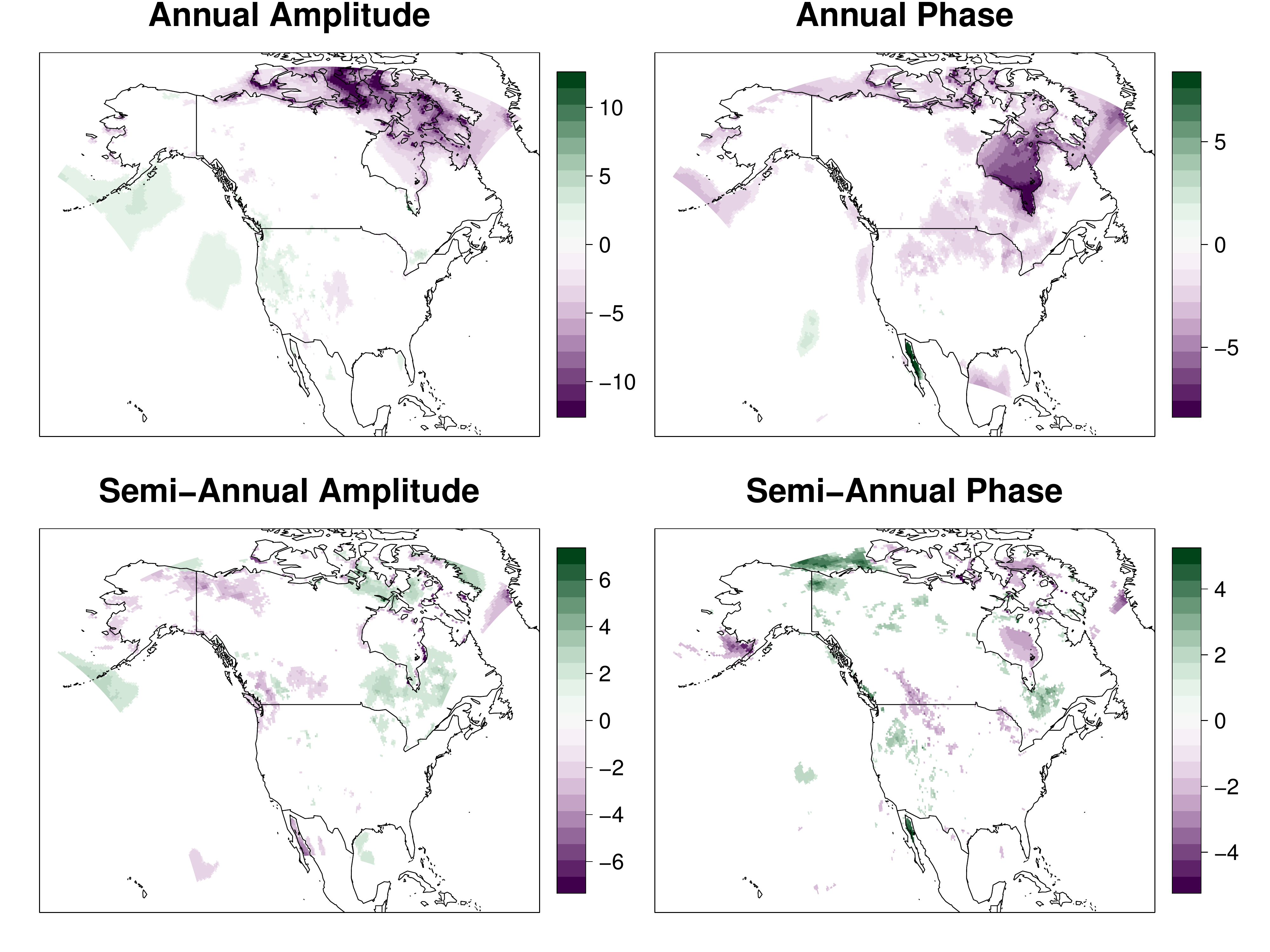}
    \caption{Computed t-statistics of the Fourier coefficients for the maximum temperature cycle. Locations with 95\% point-wise confidence intervals not including 0 are shown, with the color corresponding to the t-statistic value.}
    \label{fig:fftplots_max}
\end{figure}

%% Here, we want to investigate changes, so we focus in on two different time periods. 
To identify differential change in space through time in the annual and semi-annual harmonics, we investigate daily temperature data for each location for two separated 15-year time periods; period one from 1979 - 1994 and period two from 2003 - 2018.
From the annual and semi-annual harmonic estimates, we calculate the phase and amplitude for each year for each of the two time periods. 
As an exploratory comparison of the two periods at each location, we compute t-statistics of the difference (second period - first period) for both the annual and semi-annual phase and amplitude.
Figures \ref{fig:fftplots_min} and \ref{fig:fftplots_max} show the t-statistics over the region for all four components ($A_1$, $\varphi_1$, $A_2$, $\varphi_2$) for minimum and maximum temperature, respectively.
T-statistics that are large (in magnitude) suggest possible shifts in temperature cycles.
For example, in both the minimum and maximum temperature annual phase, a large area in the northeast of the continent has experienced a negative shift, implying the peak of the wave is occurring earlier in the recent years.
Additionally, the shift in the annual amplitude for both the minimum and maximum cycles are alike, with areas over the Pacific, central United States, and northern Canada having similar spatial patterns and values.
The shifts in the semi-annual amplitude and phase of both cycles have similar spatial patterns, with areas in the northwest United States and northern Canada most closely resembling each other.
These results for the annual amplitude and phase closely resemble the results over North America reported in \citet{Stine2009}, however our findings are purely exploratory as they do not account for any spatial, temporal, or process dependence, which could have important effects on reported regions of significant change.

%% Here's what these preliminaries motivated us to do!
These preliminary analyses identified possible shifts in the annual and semi-annual harmonic components of minimum and maximum temperature over North America, which suggest changes in the cycles themselves, and that these changes might vary considerably across space. 
Our aim is obtain full probabilistic inference with regard to these changes in minimum and maximum temperature cycles across the region. 
Specifically, we propose a multivariate statistical DSTM that captures the relationship between the minimum and maximum temperature cycles, as well as spatial and temporal dependence.  This model will be used to quantify the uncertainty associated with potential seasonal cycle changes over space and time.

%%%%%%%%%%%%%%%%%%%%%%%%%%%%%%%%%%%%%%%%%%%%%%%%%%%%%%%%%%%%%%%%%%%%%%
\ssection{Bayesian Hierarchical Formulation}
\label{sec:methods}
%%%%%%%%%%%%%%%%%%%%%%%%%%%%%%%%%%%%%%%%%%%%%%%%%%%%%%%%%%%%%%%%%%%%%%

%%%%%%%%%%%%%%%%%%%%%%%%%%%%%%%%%%%%%%%%%%%%%%%%%%%%%%%%%%%%%%%%%%%%%%
\ssubsection{Data model}

Let $\vz_{1\ell}(\vs) = [z_{11}(\vs), ..., z_{1T_{\ell}}(\vs)]'$ and $\vz_{2\ell}(\vs) = [z_{21}(\vs), ..., z_{2T_{\ell}}(\vs)]'$ denote the centered, by year, minimum and maximum temperature, respectively, for year $\ell$, $\ell = 1, \dots, L$  at location $\vs$, $\vs \in \{\vs_1, ..., \vs_n\}$ where $T_{\ell}$ is the number of days in year $\ell$. 
The linear model for temperature at location $\vs$, year $\ell$, and variable $j$ can be specified in terms of the Fourier coefficients (e.g., Eqn. \ref{eq:harm_exp}) by 
\begin{align}\label{eq:data}
    \vz_{j\ell}(\vs) = \vX_{\ell}\widetilde{\vbe}_{j\ell}(\vs) + \widetilde{\vep}_{j\ell}(\vs) \qquad j = 1,2,
\end{align}
where $\vX_{\ell} = [\vrho_1, \vpsi_1, \vrho_2, \vpsi_2]$, and $\widetilde{\vbe}_{j\ell}(\vs) = [a_1(\vs), b_1(\vs), a_2(\vs), b_2(\vs)]'$, with the $t^{th}$ element of $\vrho_h$ and $\vpsi_h$ equal to  $\rho_{ht} =\cos(2 \pi h (t-1)/T_{\ell})$ and $\psi_{ht} = \sin(2 \pi h (t-1)/T_{\ell})$, respectively, for $h=1, 2$, and  $\widetilde{\vep}_{j\ell}(\vs) \overset{iid}{\sim} \Gau(0, \widetilde{\sigma}^2_{\varepsilon_j}(\vs)\,\vI_{T_{\ell}})$. 
Here, $\widetilde{\sigma}^2_{\varepsilon_j}(\vs)$ is the variance for the $j^{th}$ variable at location $\vs$, which is assumed constant over years for each cycle and location. 
Although the simplifying assumption of iid errors assumes no residual temporal autocorrelation as would be present in ``weather'' processes, preliminary analyses that accounted for this correlation through a daily random effect did not substantially impact parameter inference.
As such, the model with daily random effects was not considered further due to the added computational complexity.

Spatial and temporal dependence is modeled using spatially-varying harmonic coefficients and with a random walk time structure.
Specifically, the harmonic coefficients, $\widetilde{\vbe}_{\ell}$, are spatial processes and the spatial field evolves according to a random walk.
Letting $p=4$ denote the number of Fourier coefficients per cycle, the $2p$-vector of spatially-varying coefficients, $\widetilde{\vbe}_{\ell}(\vs) = \left[\widetilde{\vbe}_{1\ell}(\vs)^{'}, \widetilde{\vbe}_{2\ell}(\vs)^{'}\right]^{'}$, for year $\ell$, location $\vs$ is modeled by
\begin{align}\label{eq:param_dist}
    \widetilde{\vbe}_{\ell}(\vs)|\widetilde{\vbe}_{\ell-1}(\vs),\widetilde{\vw}_{\ell}(\vs) \sim \Gau\left(\widetilde{\vbe}_{\ell-1}(\vs) + \widetilde{\vw}_{\ell}(\vs), \widetilde{\vSig}_{\beta}\right)
\end{align}
where
$\widetilde{\vw}_{\ell}(\vs) = \left[\widetilde{\vw}_{1\ell}(\vs)^{'},\widetilde{\vw}_{2\ell}(\vs)^{'}\right]^{'}$ with $\widetilde{\vw}_{j\ell}(\vs) = [\widetilde{w}_{j1\ell}(\vs), ..., \widetilde{w}_{jp\ell}(\vs)]'$, and 
\begin{align*}
    \widetilde{w}_{jk\ell}(\vs) \overset{\text{ind.}}{\sim}GP\left(0, C\left(\cdot; \vtheta_{jk}\right)\right),
\end{align*}
where $k = 1, ..., p$, $\widetilde{\vSig}_{\beta}$ is a $2p \times 2p$ unstructured covariance matrix, and $GP(0, C(\cdot; \vtheta_{jk}))$ denotes a Gaussian process over the spatial domain \citep{Cressie1992}.
Each spatial process $\widetilde{\vw}_{\ell}(\vs)$ accounts for the residual spatial variation in the Fourier coefficients between year $\ell-1$ and $\ell$.
We assume an exponential covariance function, where $C(\vs, \vs'; \vtheta_{jk}) = \sigma_{jk}^2\exp\{-||\vs - \vs'||/\phi_{jk}\}$, $||\vs - \vs'||$ is the Euclidean distance between locations $\vs$ and $\vs'$, and $\vtheta_{jk} = \{\sigma^2_{jk}, \phi_{jk}\}$ consists of the spatial variance and decay parameters, respectively, for process $k = 1, ..., p$.
Note that the spatial covariance is process and parameter specific, but assumed constant across years.

Based on the similarities in the parameter estimates discussed in Section \ref{sec:motivation}, the harmonic coefficients for both the minimum and maximum cycles are modeled jointly to borrow strength.
Dependence between the minimum and maximum temperature cycles is captured through the covariance structure in the coefficients, $\widetilde{\vSig}_{\beta}$.
Lastly, we let $\widetilde{\vbe}_0(\vs)|\vmu_0 \sim \Gau(\vmu_0, \Sigma_0)$, completing the model specification.

%%%%%%%%%%%%%%%%%%%%%%%%%%%%%%%%%%%%%%%%%%%%%%%%%%%%%%%%%%%%%%%%%%%%%%
\ssubsection{Predictive Process}\label{sub:predproc}
Model inference presents computational challenges due to the number of spatial locations, years, harmonics, and processes. For example, a single draw from the conditional distribution of $\widetilde{\vbe}_{\ell}(\vs)|\widetilde{\vbe}_{\ell-1}(\vs),\widetilde{\vw}_{\ell}(\vs)$ requires matrix operations on a $2pn \times 2pn$ matrix, which is computationally prohibitive for even modest sized data sets. Therefore, we propose using spatio-temporal predictive processes \citep{Finley2012} to enhance computational efficiency.

Let $\mathbb{S} = \{\vs_1, ..., \vs_n\}$ be the locations where minimum and maximum temperature data are available. Next, define knot locations $\mathbb{S}^{*} = \{\vs^{*}_1, ..., \vs^{*}_{m}\}$ located inside the domain of interest where $m \ll n$.  
For cycle $j$, at location $\vs$, year $\ell$, for process $k$, we define the predictive process \citep{Finley2012} as 
\begin{align}\label{eq:pred_proc}
    w_{jk\ell}(\vs) = E(\widetilde{w}_{jk\ell}(\vs)|\widetilde{\vw}_{jk\ell}^{*}) = \vc(\vs;\vtheta_{jk})'\vC^{*}(\vtheta_{jk})^{-1}\widetilde{\vw}_{jk\ell}^{*},
\end{align}
where $\widetilde{\vw}_{jk\ell}^{*} = [\widetilde{w}_{jk\ell}(\vs_1^{*}), ..., \widetilde{w}_{jk\ell}(\vs_m^{*})]^{'}$, $\vc(\vs;\vtheta_{jk})'$ is a $1 \times m$ vector whose $a^{th}$ element is $C(\vs, \vs^*_a; \vtheta_{jk})$, and  $\vC^{*}(\vtheta_{jk})$ is the $m \times m$ matrix with element $(a,b)$ given by $C(\vs^*_a, \vs^*_b; \theta_{jk})$. 
Let $\vbe_{\ell}(\vs) = \left[\vbe_{1\ell}'(\vs), \vbe_{2\ell}'(\vs)\right]^{'}$ and $\vw_{\ell}(\vs) = \left[w_{11\ell}(\vs), ..., w_{1p\ell}(\vs), w_{21\ell}(\vs), ..., w_{2p\ell}(\vs)\right]^{'}$, then the predictive process for the coefficients at year $\ell$, $\vbe_{\ell}(\vs)$, is predicated on all previous predictive processes,
\begin{align*}
    \vbe_{\ell}(\vs) = \sum_{r=1}^{\ell}\vw_r(\vs) + \eta_r,
\end{align*}
%\begin{align}
%    \vbe_{\ell}(\vs) & \sim \Gau\left(\sum_{r=1}^{\ell}\vw_r(\vs) + \eta_r, \vSig_{\beta}\right),
%\end{align}
where $\eta_r \sim \Gau(0, \vSig_{\beta})$ and $\vSig_{\beta}$ is defined as in Eqn. \ref{eq:param_dist}.  The resulting distribution of the coefficients is
\begin{align*}
    \vbe_{\ell}(\vs)|\vbe_{\ell-1}(\vs), \vw_{\ell}(\vs) \sim \Gau\left(\vbe_{\ell-1}(\vs) + \vw_{\ell}(\vs), \vSig_{\beta}\right),
\end{align*}
and the data model (Eqn. \ref{eq:data}) can be rewritten in the form of the predictive process.
\begin{align*}
    \vz_{j\ell}(\vs) = \vX_{\ell}\vbe_{j{\ell}}(\vs) + \vep_{j{\ell}}(\vs), \qquad j = 1,2,
\end{align*}
where $\vep_{j\ell}(\vs) \overset{iid}{\sim} \Gau(0, \sigma^2_{\varepsilon_j}(\vs)\, \vI_{T_{\ell}})$ is defined the same as in Eqn. \ref{eq:data}.
Therefore, conditioned on the predictive process, the coefficient process is spatially independent, and draws from the full conditional of $\vbe_{\ell}(\vs)$ can be obtained univariately.

\ssubsection{Parameter Models}

To fully specify the Bayesian hierarchical model, we assign prior distributions to all remaining parameters.
Conjugate, non-informative priors were chosen when available to ease computational burden.
For $\sigma^2_{\epsilon_{\ell}}(\vs)$, the variance for location $\vs$ that is shared across time is modeled $\sigma^2_{\epsilon_j}(\vs)  \sim \text{Inv-Gamma}(a,b)$.
For the $2p \times 2p$ covariance matrix of the $\beta$ parameters we assign $\vSig_{\beta} \sim \text{Inv-Wishart}(\vV, \xi)$.
Lastly, the spatial variance for the $k^{th}$ spatial process is modeled $\sigma_{jk}^2  \sim \text{Inv-Gamma}(a_{jk}, b_{jk})$.
All hyperpriors were chosen such that the priors have finite first moments, specifically $\vV = \vI_8$, $\xi = 11$, and $a = b = a_k = b_k = 2$ for all $k$.
Preliminary analyses with a uniform prior distribution for the spatial decay parameter, $\phi_{jk}$ indicated that this parameter had little impact on the inference of the parameters of interest. Therefore, we set $\phi_{jk} = \phi$ to a fixed value in our analysis.
%Additional details on this specification are provided in Section \ref{sec:results}.

The full hierarchical model can be written
\begin{align}\label{full_mod}
\begin{aligned}
    & \prod_{s=1}^n\prod_{\ell=1}^L\prod_{j=1}^2
    \Big[\vz_{j\ell}(\vs)|\vbe_{j\ell}(\vs), \sigma^2_{\varepsilon_j}(\vs)\Big]
    \prod_{s=1}^n\prod_{\ell=1}^L
    \Big[\vbe_{\ell}(\vs) | \vbe_{\ell-1}(\vs), \vw_{\ell}(\vs), \vSig_{\beta}, \vtheta_{jk} \Big] \\
    & \prod_{s=1}^n
    \Big[ \vbe_0(\vs) | \vmu_0, \vSig_0 \Big]
    \prod_{s=1}^n\prod_{\ell=1}^L
    \Big[ \widetilde{\vw}_{\ell}^{*}(\vs)|\vtheta_{jk} \Big] 
    \Big[ \vSig_{\beta} \Big]
    \prod_{s=1}^n\prod_{j=1}^2
    \Big[ \sigma^2_{\epsilon_j}(\vs) \Big]
    \prod_{j=1}^2\prod_{k=1}^p
    \Big[\sigma^2_{jk}\Big],
\end{aligned}
\end{align}
where $\vw_{\ell}(\vs)$ is a deterministic composition of $\widetilde{\vw}_{\ell}^*$, as shown in Eqn. \ref{eq:pred_proc}.

\ssubsection{Model Inference}

Recall from Section \ref{sec:motivation} that the motivation for this modeling effort is purely inferential. Specifically, we are interested in inference with respect to the spatial processes of harmonic coefficients, $\vbe_{\ell}(\vs)$ for $\ell \in \{1, ..., L\}$. We obtain samples from the joint posterior distribution using a Gibbs sampling algorithm, the details of which are given in Appendix \ref{sec:gibbs_sample}.  Each of the parameters  described above have a conjugate full-conditional distributions.
To improve computational efficiency of our sampling algorithm, we also took advantage of parallel computation when possible.  Specifically, conditioned on the predictive process, $\vw_{\ell}$, the parameters $\vbe_{\ell}$ are spatially independent and can be updated in parallel.  

Posterior inference will focus on the comparison of the amplitude and phase of the annual and semi-annual harmonics across all years and spatial locations. 
Visualizing their spatial evolution over time provides insight into how the temperature cycle changes across years. 
Using samples from the posterior distribution of $\vbe_{\ell}(\vs)$, we can obtain full posterior inference for the phase and amplitude of the annual and semi-annual harmonics using composition sampling. We can also compute important characteristics of these cycles, such as the day at which the cycle reached its peak or trough. These peak and trough days can be compared across time to quantify shifts in temperature cycles which may have important impacts on spatial synchrony. 
In addition, we can identify similarities and differences in shifts in minimum and maximum temperature cycles.

%%%%%%%%%%%%%%%%%%%%%%%%%%%%%%%%%%%%%%%%%%%%%%%%%%%%%%%%%%%%%%%%%%%%%%
\ssection{Results}
\label{sec:results}
%%%%%%%%%%%%%%%%%%%%%%%%%%%%%%%%%%%%%%%%%%%%%%%%%%%%%%%%%%%%%%%%%%%%%%
We fitted the model to the NCEP Reanalysis temperature data introduced in Section \ref{sec:motivation}.
The spatial domain of interest spanned the continental United States and portions of Mexico and Canada, with corners at approximately (25.039N, 120.243W), (21.399N, 79.588W), (46.471N, 64.321W), and (52.7418N, 128.744W).  
We thinned the data spatially to reduce the overall dimension, keeping every other location in both the longitudinal and latitudinal directions. This resulted in daily minimum and maximum temperature values at 3621 spatial locations over 40 years, for a total of over $1.06\times 10^{8}$ data points. 
For the predictive process outlined in Section \ref{sub:predproc}, we chose 144 knot locations evenly spaced across the domain of interest.
All temperature time series were centered since the focus of inference is on the harmonics and change in harmonics as opposed to raw temperature.

Using MCMC and the Gibbs sampling algorithm (see Appendix \ref{sec:gibbs_sample}), we obtained 5000 samples from the joint posterior distribution. 
The first 1000 samples were discarded as burn-in and the remaining 4000 samples were retained for posterior inference.
All computation and posterior inference was performed on a high performance computing infrastructure\footnote{Computation was performed on a Linux workstation using an Intel(R) Xeon(R) CPU E5-2680 v4 @ 2.40GHz processor utilizing 24 cores.} due to the dimensionality of the data and Bayesian inference output.
Convergence of model parameters was assessed visually via trace plots, with no issues detected.  

\begin{figure}[ht]
    \centering
    \includegraphics[width = \linewidth]{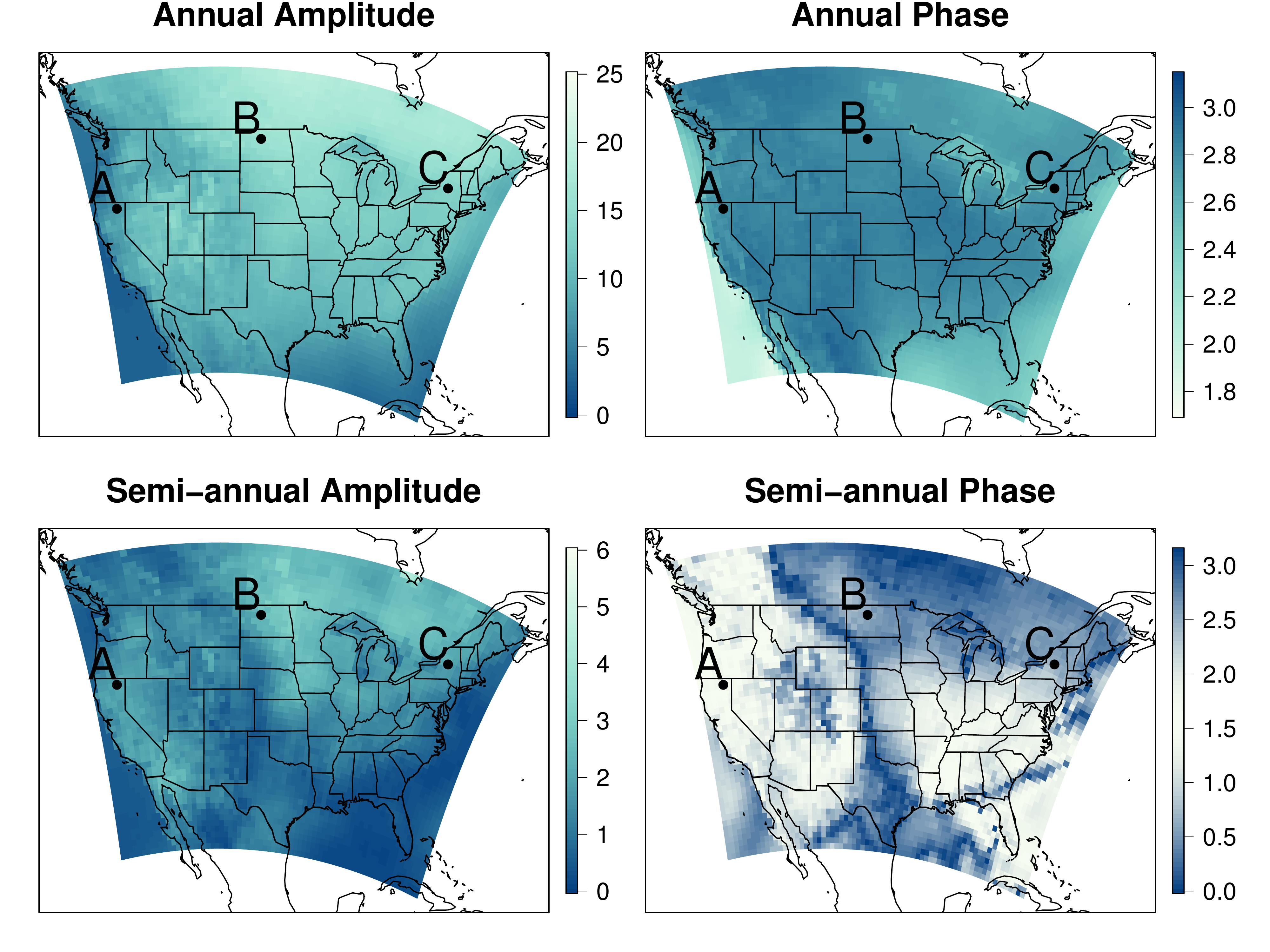}
    \caption{Posterior mean estimates of the minimum temperature annual and semi-annual harmonics for 2004.  The left two panels show the annual and semi-annual amplitude (deg C), and the right tow panels show the annual and semi-annual phase (radians).  Location A corresponds to the top plot in Figure \ref{fig:directional_wave}; Location B corresponds to the middle plot in Figure \ref{fig:directional_wave}; Location C corresponds to the bottom plot in Figure \ref{fig:directional_wave}.}
    \label{fig:animation_snap_min}
\end{figure}

\begin{figure}[t]
    \centering
    \includegraphics[width = \linewidth]{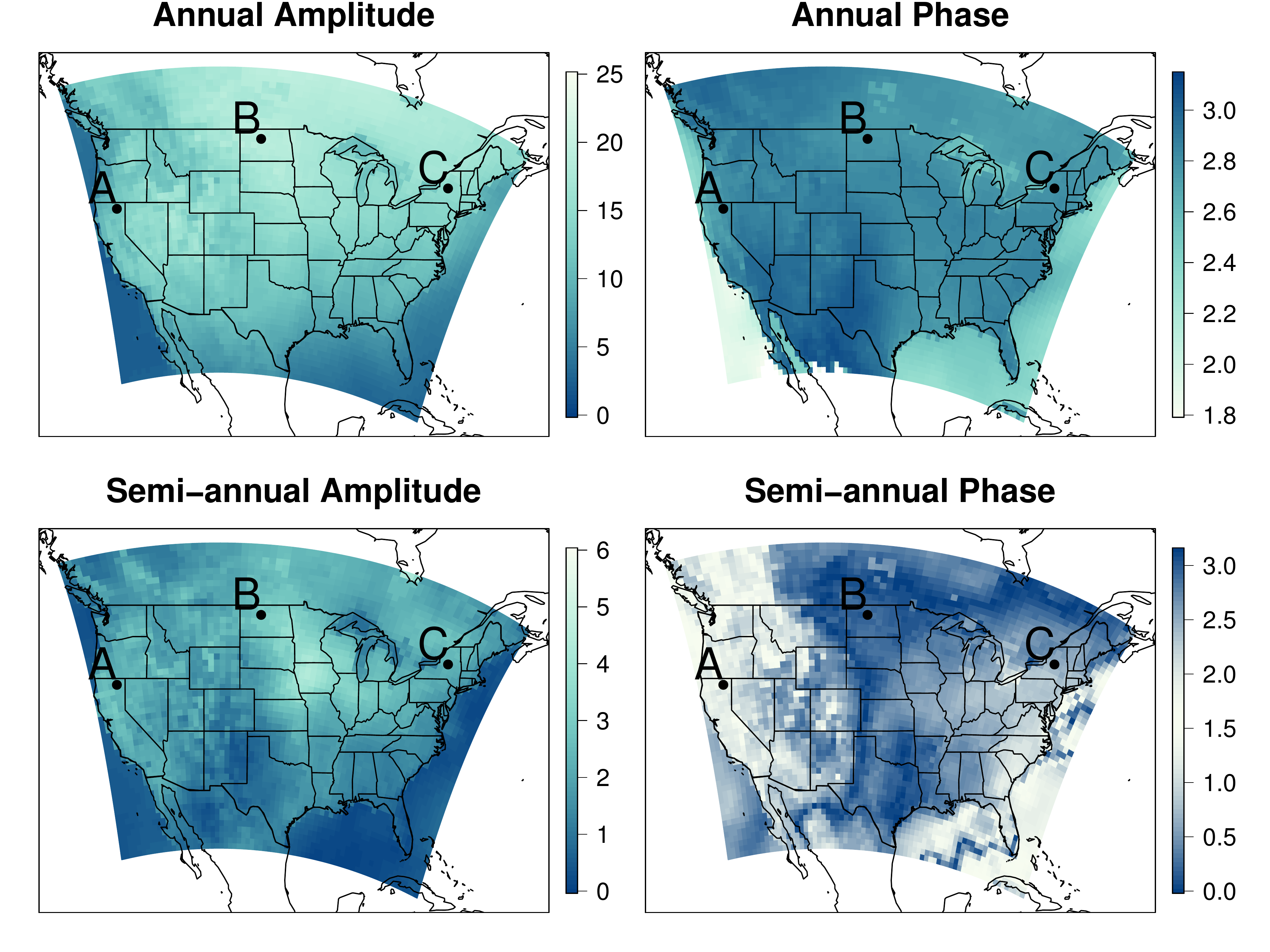}
    \caption{Posterior mean estimates of the maximum temperature annual and semi-annual harmonics for 2004.  The left two panels show the annual and semi-annual amplitude (deg C), and the right tow panels show the annual and semi-annual phase (radians).  Location A corresponds to the top plot in Figure \ref{fig:directional_wave}; Location B corresponds to the middle plot in Figure \ref{fig:directional_wave}; Location C corresponds to the bottom plot in Figure \ref{fig:directional_wave}.}
    \label{fig:animation_snap_max}
\end{figure}

Posterior distributions of the annual and semi-annual phase and amplitude of minimum and maximum temperature cycles were obtained for each year using composition sampling. 
Posterior mean estimates of these four cycle quantities for minimum and maximum temperature for the year 2004 are shown in Figures \ref{fig:animation_snap_min} and \ref{fig:animation_snap_max}, respectively.
For both figures, estimates of the annual and semi-annual harmonics are shown on the top and bottom panels, respectively, while the amplitude and phase estimates are shown on the left and right, respectively. 
A prominent spatial feature of the temperature harmonics is the wave-like pattern that appears in the semi-annual amplitude and phase.
This wave can been seen most prominently in the bottom-right panel of Figure \ref{fig:animation_snap_min} where a band of semi-annual phase values close to 0 (or $\pi$) spans from the north-west United States down through the center of the United States. 
The semi-annual phase estimates on either side of this band deviate from 0 (or $\pi$).
The same spatial pattern appears in the semi-annual amplitude, where there is a band of smaller amplitudes following approximately the same path.

These same spatial patterns can are seen in Figure \ref{fig:animation_snap_max} for the semi-annual components of the maximum temperature cycle.
In the bottom-right panel, the western United States have a contiguous area of lighter colored values close to 1.5 that are bordered by values close to 0 (or $\pi$).
While less prominent than the minimum semi-annual amplitude, the maximum semi-annual amplitude has a band of smaller amplitudes that spans from the southern United States up through the center of the United States.
These spatial patterns likely arise because of the east-west structure of the semi-annual harmonic due to the land/sea contrast as discussed in the introduction and by \citet{Wikle1996}.

To investigate the temporal variation in minimum and maximum temperature cycles throughout the 40 year period, we produced an animation of the posterior mean estimates of amplitude and phase for the first and second harmonics\footnote{\href{https://joshuanorth.shinyapps.io/harmonics_application/}{https://joshuanorth.shinyapps.io/harmonics\_application/}}.
The animation illustrates the slow evolution of the annual components and the temporal volatility of the semi-annual components.
While the wave-like patterns seen in 2004 (Figures \ref{fig:animation_snap_min} and \ref{fig:animation_snap_max}) are the most common, variations of these 

\begin{figure}
    \centering
    \includegraphics[width = 0.95\linewidth]{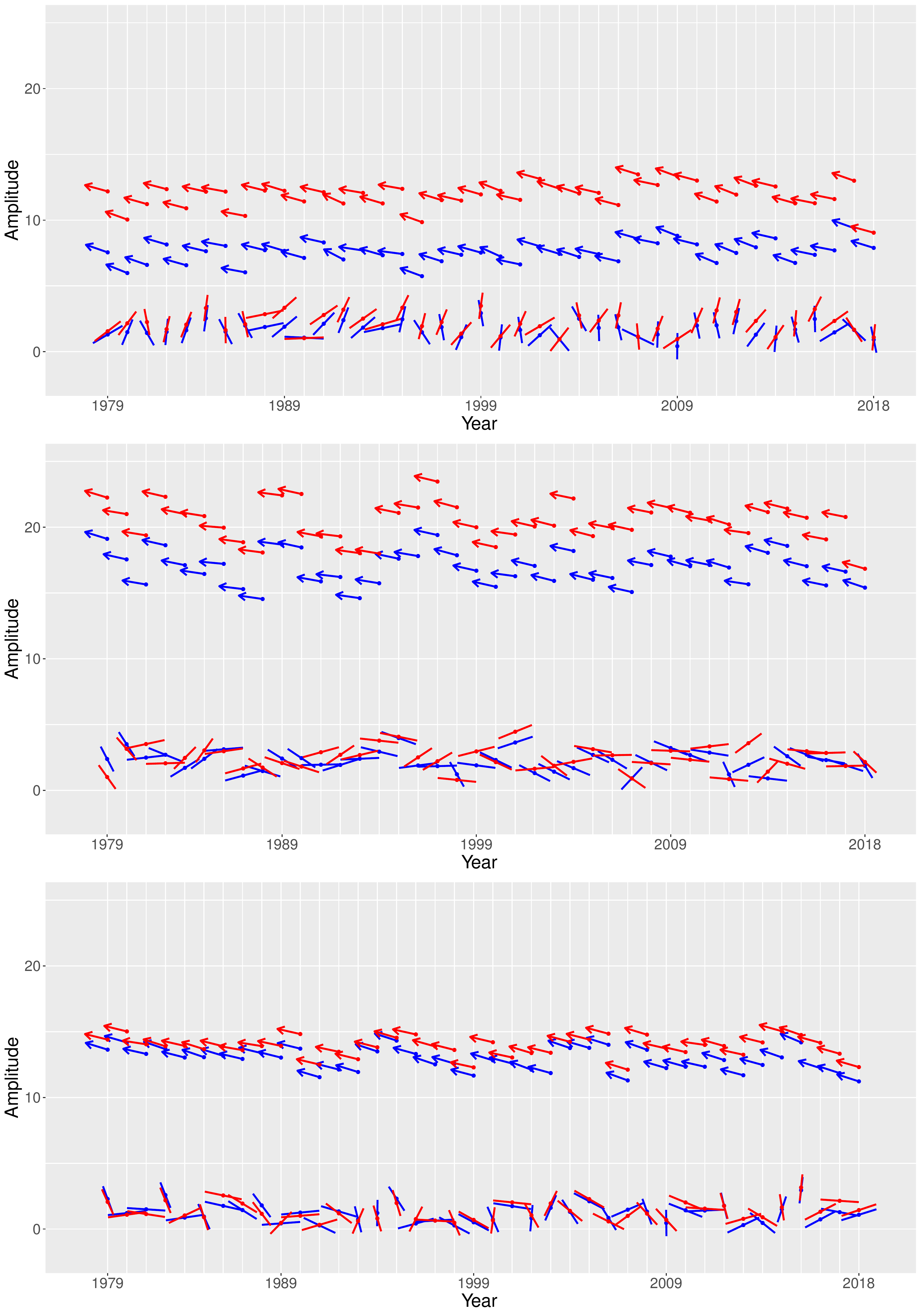}
    \caption{The phase in radians (angular direction of the arrow/line), $\varphi \in [0, 2\pi]$, and amplitude (height of the point) in degrees Celsius, $A \in [0, \infty)$, for the minimum (blue) and maximum (red) temperature cycles at three spatial locations across the United States.  See Figure \ref{fig:animation_snap_min} corresponding to geographic locations, with the top plot corresponding to location A, middle plot to location B, and bottom plot to location C.}
    \label{fig:directional_wave}
\end{figure}
\FloatBarrier

\noindent spatial patterns appear in both the maximum and minimum semi-annual amplitude and phase in other years.
Similar wave-like spatial patterns have been detected for geopotential height \citep{Wallace1993, Thompson1998, Thiebaux1986, Wikle1996} as discussed in the introduction.

To illustrate the component-wise difference in minimum and maximum temperature cycles, Figure \ref{fig:directional_wave} shows posterior estimates of the annual and semi-annual amplitudes (height of the point on the y-axis) and phase (angular direction of the arrow) simultaneously for minimum and maximum temperature at three different spatial locations.
The locations, denoted ``A", ``B", and ``C" in  Figures \ref{fig:animation_snap_min} and \ref{fig:animation_snap_max}, are in the west, north central, and northeast, respectively.
The relationship between the minimum and maximum annual amplitudes differ across space, which could be attributed to climatological variations.
Specifically, for locations ``A'' and ``B'', the range between the minimum and maximum annual amplitudes is greater than for location ``C'', and the interannual variability for the annual amplitudes is much greater for location ``B'' than locations ``A'' and ``C''.
The minimum and maximum semi-annual phase are the same for most years (i.e., phase locked), with relatively little year-to-year change in the semi-annual amplitude.
Compared to the annual phase, the semi-annual phase is more volatile with each location experiencing differing degrees of variability. The semi-annual phase appears the most variable for location ``C''.

The extent to which the temperature cycles have shifted (i.e., how the temperature cycle determined by the estimates of the first two harmonics has changed over time) over the 40 year period can be seen by comparing the day of the year at which the temperature cycles are at their peak and trough.  We computed the average peak and trough day for the years 1979-1988 and 2009-2018 as well as the differences between these two time periods (computed as 2009-2018 minus 1979-1988).
These posterior distributions can be used to identify spatial regions experiencing significant shifts in the minimum and maximum temperature cycles.  We consider a shift to be significant if the 95\% credible interval of the difference does not include zero.

\begin{figure}[ht]
    \centering
    \includegraphics[width = \linewidth]{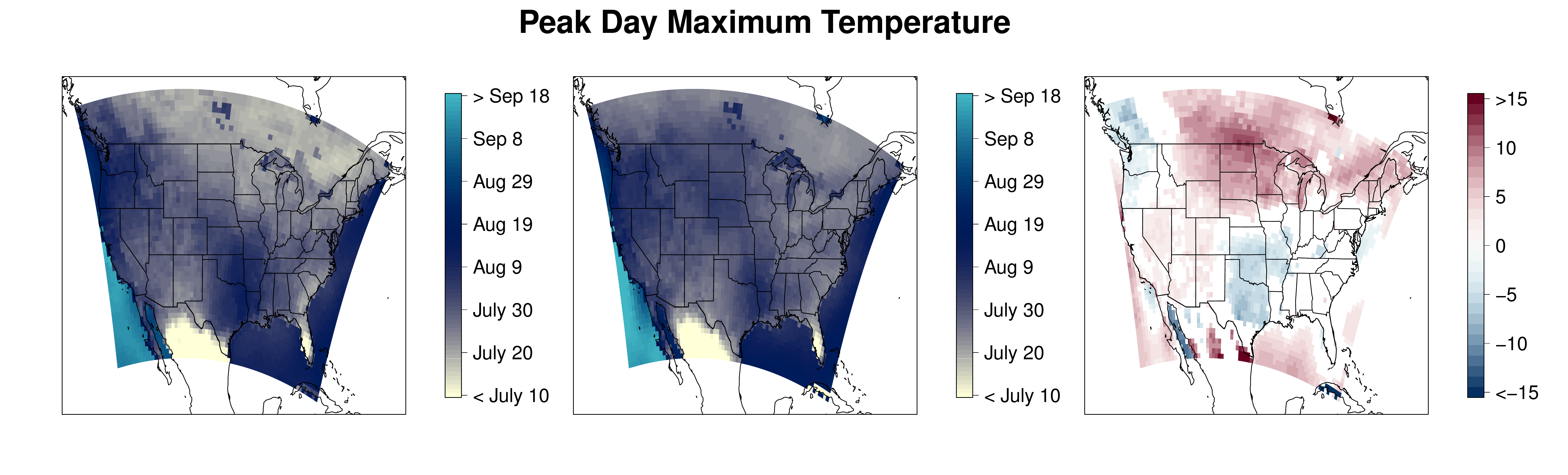}
    \caption{Day at which the maximum temperature cycle reaches its peak.  The left image is the average over the years 1979-1988, the center image is the average over the years 2009-2018, and the right image is the difference between the two images (2009-2018 minus 1979-1988), showing only locations where there is a significance difference.}
    \label{fig:max_peak}
\end{figure}

\begin{figure}[ht]
    \centering
    \includegraphics[width = \linewidth]{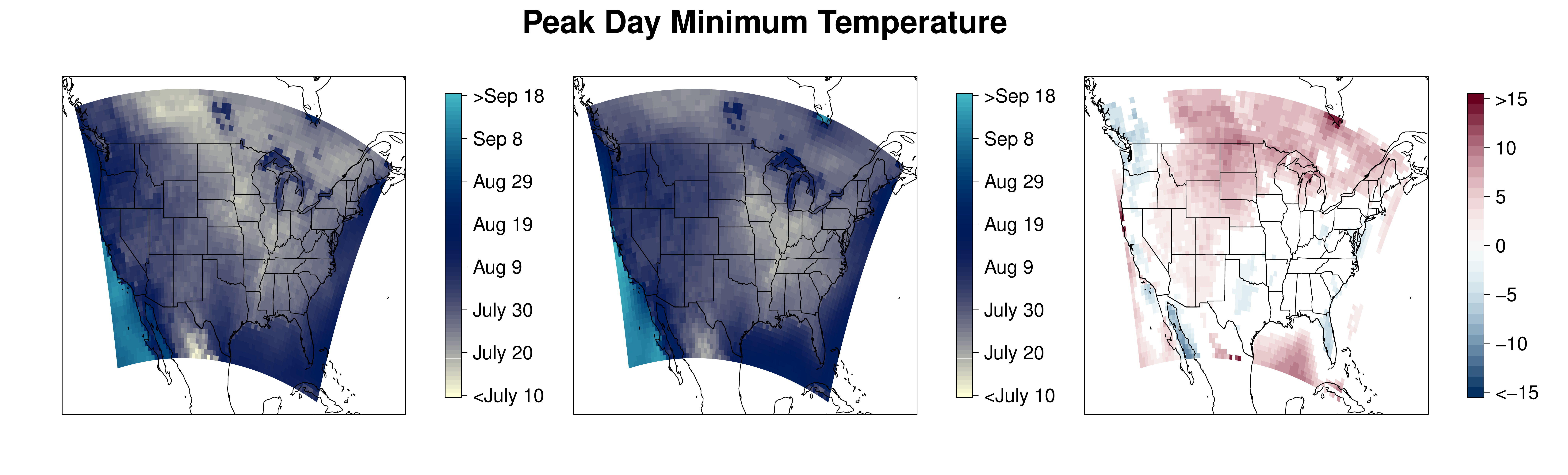}
    \caption{Day at which the minimum temperature wave reaches its peak. The left image is the average over the years 1979-1988, the center image is the average over the years 2009-2018, and the right image is the difference between the two images (2009-2018 minus 1979-1988), showing only locations where there is a significance difference.}
    \label{fig:min_peak}
\end{figure}

Figures \ref{fig:max_peak} and \ref{fig:min_peak} show the average day of the year in which the maximum and minimum temperature cycles, respectively, obtain their peak, as determined by the first two harmonics. The peak day for maximum temperature can be thought of as the hottest day of the year, and the peak day for minimum temperature is the day at which the warmest low temperature occurs.  The differences in peak days between the two decades for both maximum and minimum temperature clearly identify regions experiencing seasonal shifts in temperature.  The areas in red indicate locations for which the peak day is occurring later in the year for the 2009-2018 decade, whereas blue regions correspond to locations in which the peak day is occurring earlier in the year for the more recent decade. The spatial patterns in the shifts in maximum and minimum temperature appear similar across the domain. The northern regions (Montana, the Dakotas, Minnesota, and Canada) appear to be experiencing the greatest shift towards later seasonal peaks, with much of the western United States experiencing more moderate shifts.  For both minimum and maximum temperature, the only two areas experiencing a shift to earlier seasonal peaks are located in the Midwest United States and along the Northwest coast.

\begin{figure}[ht]
    \centering
    \includegraphics[width = \linewidth]{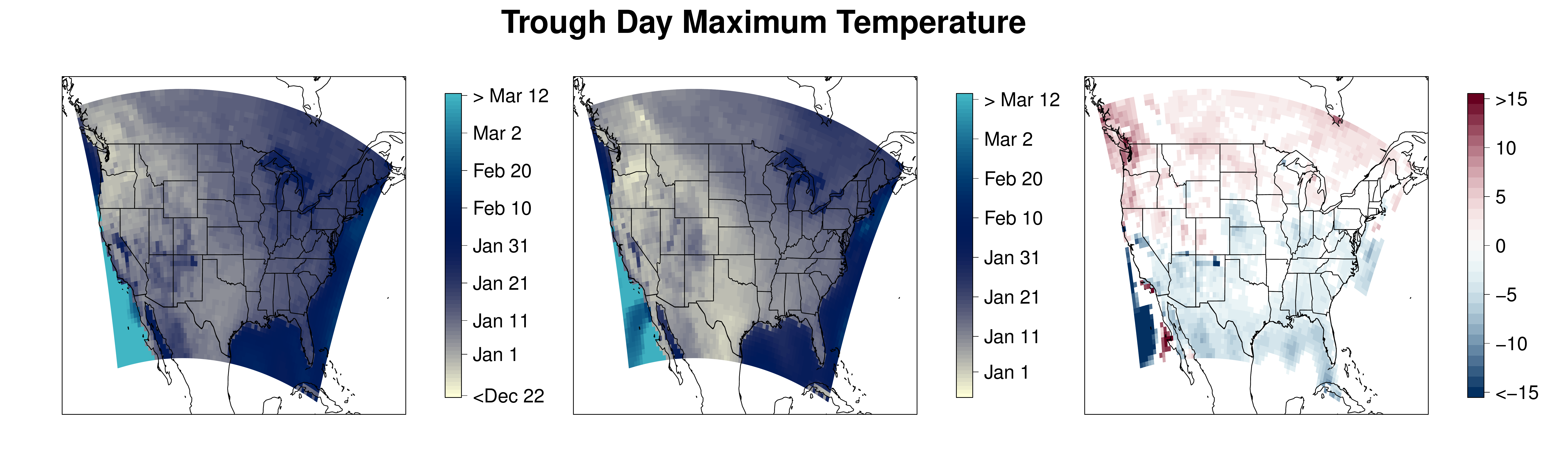}
    \caption{Day at which the maximum temperature cycle reaches its lowest. The left image is the average over the years 1979-1988, the center image is the average over the years 2009-2018, and the right image is the difference between the two images (2009-2018 minus 1979-1988), showing only locations where there is a significance difference.}
    \label{fig:max_trough}
\end{figure}

\begin{figure}[t]
    \centering
    \includegraphics[width = \linewidth]{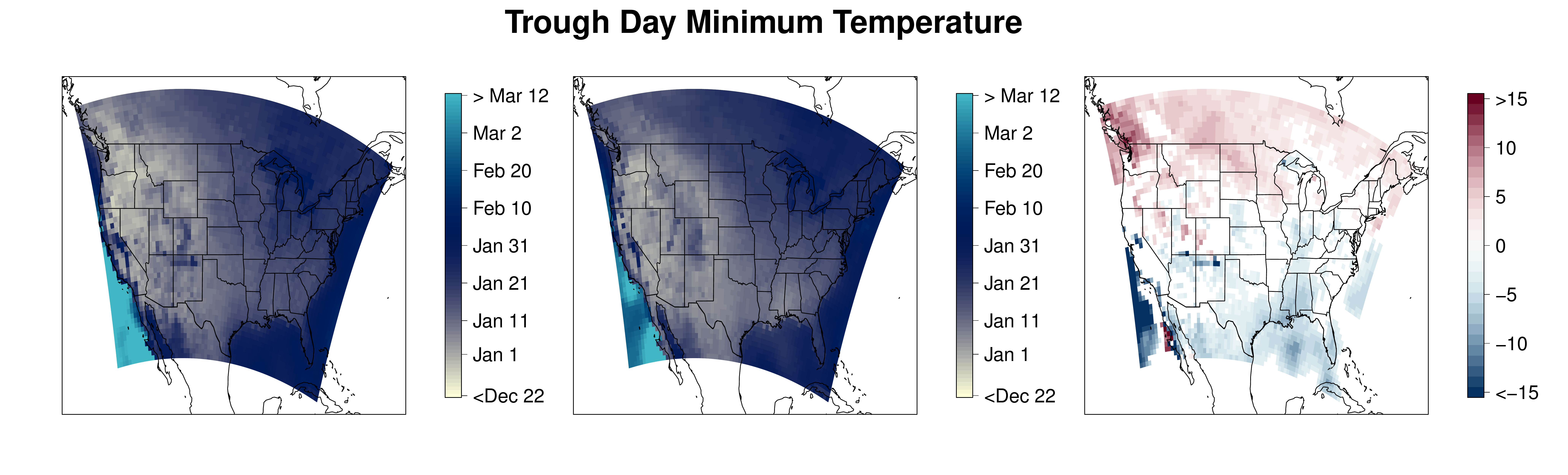}
    \caption{Day at which the minimum temperature cycle reaches its lowest. The left image is the average over the years 1979-1988, the center image is the average over the years 2009-2018, and the right image is the difference between the two images (2009-2018 minus 1979-1988), showing only locations where there is a significance difference.}
    \label{fig:min_trough}
\end{figure}

Figures \ref{fig:max_trough} and \ref{fig:min_trough} show the day for which each maximum and minimum temperature cycle reaches its trough as determined by the first two harmonics. 
For minimum temperature, the trough corresponds to the coldest day of the year, and for the maximum temperature, the trough captures the day of the coldest high temperature.  
Again, we see very similar patterns between minimum and maximum temperature cycles and shifts.
In contrast to the spatial distribution of the shift for the peak day, the spatial distribution of the shift for the trough day has a strong north/south pattern.
The northern half of the United States and Canada are experiencing a shift toward later seasonal troughs, whereas the southern half of the United States and Mexico are experiencing a shift toward earlier seasonal troughs.

\begin{figure}[ht]
    \centering
    \includegraphics[width = \linewidth]{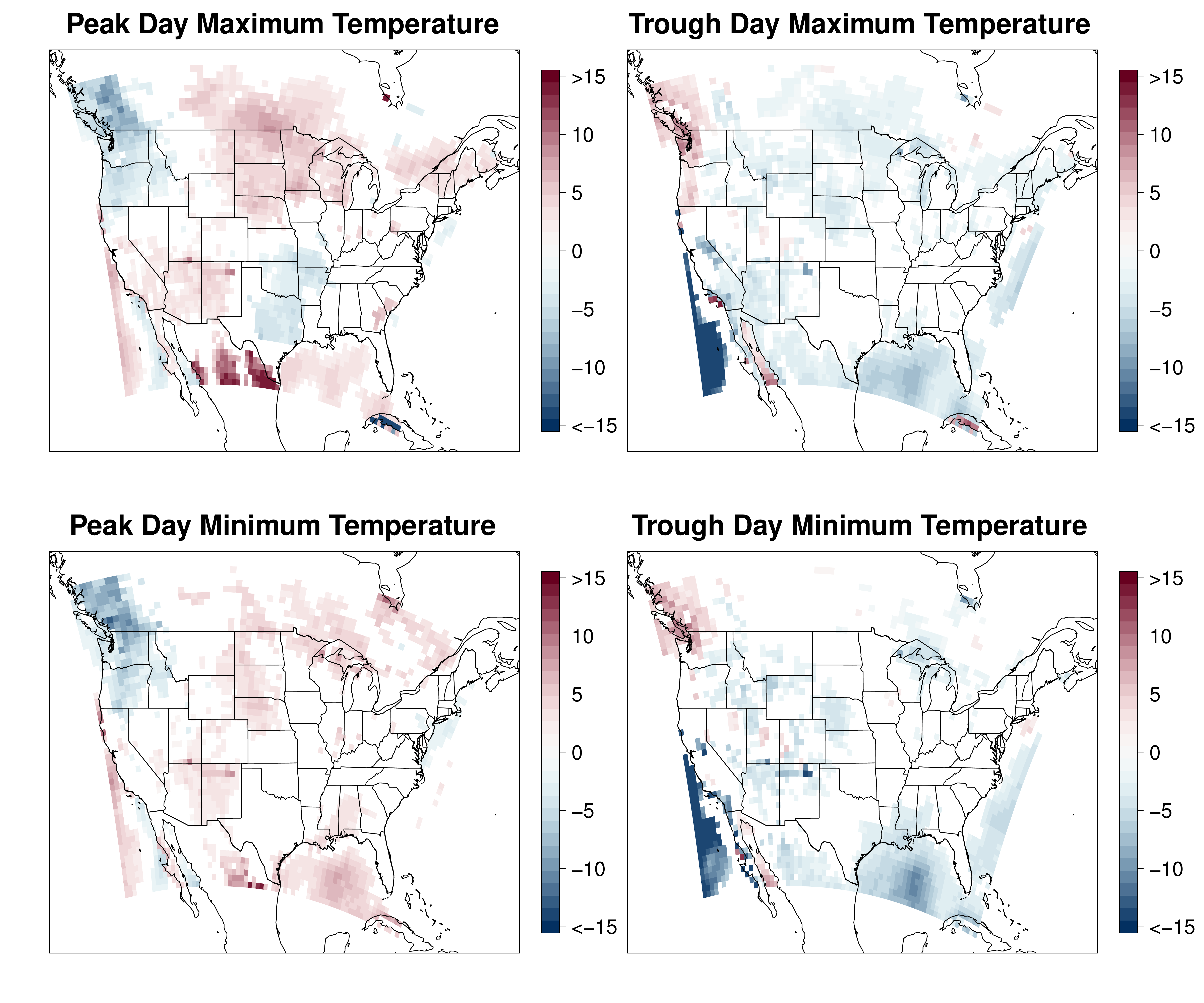}
    \caption{Difference of the decadal shifts for the estimates considering the annual and semi-annual component with estimates considering only the annual component.  Shading corresponds to the magnitude of the differences, with white areas corresponding to locations where the semi-annual component is not significant.  Red indicates the peak/trough day has occurred later in the year when the semi-annual harmonic is considered, and blue indicates the peak/trough has occurred earlier in the year.}
    \label{fig:diff}
\end{figure}

To highlight the contribution of the semi-annual component on the temperature shift, we obtained posterior distributions of the peak and trough days as well as the decadal shifts for both the minimum and maximum temperature cycles using only the annual components.
We then computed the posterior difference in the decadal shifts between those obtained when both the annual and semi-annual component were included and those when only the annual component was included. 
The posterior mean of these differences are shown in Figure \ref{fig:diff}. We considered the contribution of the semi-annual component to be significant if the 95\% point-wise credible interval of the posterior distribution of differences did not include zero.  
Based on these posterior credible intervals, locations in white correspond to locations where the semi-annual harmonic did not significantly contribute to the shift in the temperature cycle.  
All other locations indicate that the semi-annual component contributed significantly in capturing shifts in temperature cycles between the two decades.  
In the left panels of Figure \ref{fig:diff}, red (blue) indicates locations for which the peak day has shifted to later (earlier) in the year when the semi-annual harmonic is considered.
Similarly, in the right panels of Figure \ref{fig:diff}, red (blue) indicate locations where the trough day occurs later (earlier) in the year when the semi-annual harmonic is considered.
In each of these figures, the shading corresponds to the magnitude of these differences.
The spatial distribution of significant semi-annual harmonic contributions are similar for both the minimum and maximum temperature cycles. The magnitude of the shift is higher for maximum temperature than for minimum, which could be attributed to the maximum temperature having more seasonal variation than the minimum.
The semi-annual component significantly contributes to the later peak day (positive shift) in both the minimum and maximum temperature cycle in the North (North Dakota, South Dakota, and Minnesota), Southwest (New Mexico and Arizona), and the Gulf of Mexico.
The semi-annual component significantly contributes to the earlier trough day (negative shift) in both minimum and maximum temperature cycles in the Gulf of Mexico and western United States.

%%%%%%%%%%%%%%%%%%%%%%%%%%%%%%%%%%%%%%%%%%%%%%%%%%%%%%%%%%%%%%%%%%%%%%
\ssection{Discussion and Future Work}
\label{sec:discussion}
%%%%%%%%%%%%%%%%%%%%%%%%%%%%%%%%%%%%%%%%%%%%%%%%%%%%%%%%%%%%%%%%%%%%%%

We proposed modeling minimum and maximum temperature cycles jointly through the components of the annual and semi-annual harmonics using a DSTM to detect temporal changes in the seasonal temperature cycle that may vary across space.
Implementing our model in a Bayesian paradigm, we obtain estimates of the annual and semi-annual phase and amplitude through composition sampling.
Spatial maps showing the difference in peak/trough days of the minimum and maximum temperature cycles for the years 2009-2018 relative to 1979-1988 identified regions experiencing seasonal shifts, as well as regions for which the semi-annual component contributed significantly to these shifts. 
These maps showed that the peak day for both minimum and maximum temperature cycles has shifted to later in the year in northern regions, and the trough day has shifted toward later in the year in the northern half and earlier in the year in the southern half of the United States.

The results of our analysis can be compared to those presented in previous research. 
For example, a similar east-west structure in the semi-annual cycle was reported by \citet{Wikle1996} and attributed to the land-sea contrast.
Additionally, using only the annual harmonic, \citet{Stine2009} found an asymmetrical spatial pattern in the shift of the temperature cycle. However, since we considered both the annual and semiannual harmonic, our results differed from theirs in terms of the regions identified as experiencing asymmetric shifts in temperature cycles.
Lastly, our model detected spatially-varying shifts in the peak/trough of the temperature cycle ranging between 15 days earlier to 15 days later in the year, whereas \citet{Dwyer2012} reported the annual phase in the temperature cycle is shifting to only later in the year.

%As mentioned in Section \ref{sec:methods}, local weather patterns can result in small-scale spatial and temporal correlation not accounted for in the model. This additional dependence can be modeled using location-specific daily random effects. We assessed the effect of the additional temporal correlation on our model parameters and found no significant differences in the posterior distributions of the harmonic coefficient parameters. Therefore, we did not further explore this model specification due to the dramatic increase in computation time required to include these daily random effects at the spatial and temporal scale of interest.

While the results of our model have scientific merit of their own, they can also be used to detect changes in spatial synchrony between temperature cycles and other important environmental processes.
For example, incorporating estimates of shifting temperature cycles in models for bird migration could identify regions for which the spatial synchrony between migration patterns of birds and temperature cycles have been disrupted. 
Similarly, we can investigate the effects temperature shifts on the occupancy or abundance of native bark beetles, which could lead to improved predictions of beetle spread or risk of invasion as well as aid in conservation efforts.
While these are just two brief examples, a better understanding of the direction and magnitude in the shifts in temperature cycles over the last 40 years will motivate future scientific hypotheses with regard to the effects of these changes on important environmental processes.

%%%%%%%%%%%%%%%%%%%%%%%%%%%%%%%%%%%%%%%%%%%%%%%%%%%%%%%%%%%%%%%%%%%%%%
\setstretch{1}
%\bibliographystyle{apacite}
%\bibliography{references}
\bibliography{references_clean}
%%%%%%%%%%%%%%%%%%%%%%%%%%%%%%%%%%%%%%%%%%%%%%%%%%%%%%%%%%%%%%%%%%%%%%

%%%%%%%%%%%%%%%%%%%%%%%%%%%%%%%%%%%%%%%%%%%%%%%%%%%%%%%%%%%%%%%%%%%%%%
\appendix
%%%%%%%%%%%%%%%%%%%%%%%%%%%%%%%%%%%%%%%%%%%%%%%%%%%%%%%%%%%%%%%%%%%%%%
\ssection{Gibbs sampling algorithm}
\label{sec:gibbs_sample}

All code can be found at \href{https://github.com/jsnowynorth/Harmonics}{https://github.com/jsnowynorth/Harmonics}.  This includes code for downloading and processing the data, the sampler (listed below), and processing the results.

Defining notation that will be used and restructuring equations to match those used in the sampler, let
\begin{align*}
    \vz_{\ell}(\vs) & = [\vz_{1\ell}(\vs)', \vz_{2\ell}(\vs)']' \sim Gau \left( \vX_{\ell} \vbe_{\ell}(\vs), \vSig_{\varepsilon}(\vs) \right), \quad \text{where} \\
    \mathbb{X}_{\ell} & = \begin{bmatrix}\vX_{\ell} & 0 \\ 0 & \vX_{\ell} \end{bmatrix}, \quad \text{and} \quad \vSig_{\varepsilon_{\ell}}(\vs) = \begin{bmatrix}\sigma^2_{\varepsilon_1}(\vs)I_{T_{\ell}} & 0 \\ 0 & \sigma^2_{\varepsilon_2}(\vs)I_{T_{\ell}} \end{bmatrix}.
\end{align*}
Writing the $\vbe$'s in block notation, which will be used in the update of the predictive process, let $\vB_{\ell} = [\vbe_{\ell}(\vs_1)', ..., \vbe_{\ell}(\vs_n)']'$ and $\vSig_{B} = I_n \otimes \vSig_{\beta}$, resulting in
\begin{align*}
    \vB_{\ell} \sim Gau \left( \vB_{\ell-1} + \vF_{\ell} \widetilde{\vw}_{\ell}^*, \vSig_B \right),
\end{align*}
where $\vF_{\ell} = $ Block-Diag$[\vC(\theta_{11})\vC^*(\theta_{11}), ..., \vC(\theta_{2p})\vC^*(\theta_{2p})]$.

The Gibbs sampler algorithm for Eqn. \ref{full_mod} is initialized by setting all parameters to some starting values.  Then, for each iteration of the Gibbs sampler, parameters are updated:

\begin{enumerate}

    \item For ${\ell} = 1, ..., L$, update $[\widetilde{\vw}^*|\cdot] \sim Gau\left(\vV_w^{-1} \va_w, \vV_w^{-1}\right)$ where
    \begin{align*}
        \va_w & = \vF_{\ell}'\vSig_B^{-1}\left(\vB_{\ell} - \vB_{\ell-1}\right), \\
        \vV_w & = \vF_{\ell}'\vSig_B^{-1}\vF_{\ell} + \vSig_W^{-1},
    \end{align*}
    where $\vSig_W^{-1} = $ Block-Diag$[\vC^{*}(\vtheta_{11})^{-1}, ..., \vC^{*}(\vtheta_{2p})^{-1}]$.
    
    \item For $k = 1, ..., p$ and $j = 1,2$, update $[\sigma_{jk}^2|\cdot] \sim IG(a, b)$, where
    \begin{align*}
        a = \frac{Lm}{2} + a_{jk} \quad \text{and} \quad b = b_{jk} + \frac{1}{2}\sum_{\ell=1}^{L}\widetilde{\vw}_{jk\ell}^{*'} \vC^{*}(\vtheta_{jk})^{-1}\widetilde{\vw}_{jk\ell}^*,
    \end{align*}
    where $a_{jk}$ and $b+{jk}$ are chosen hyperpriors, and $\widetilde{\vw}_{jk\ell}^*$ denotes the $k^{th}$ predictive process for cycle $j$ for all knot locations.
    
    \item For $\vs = \vs_1, ..., \vs_n$, update$[\vbe_0(\vs_i)|\cdot] \sim Gau\left(\vV_0^{-1} \va_0, \vV_0^{-1}\right)$ where
    \begin{align*}
        \va_0 & = \vSig_{\beta}^{-1}(\vbe_{1}(\vs_i) - \vw_{1}(\vs_i)) + \vSig_0^{-1}\vmu_0, \\
        \vV_0 & = \vSig_{\beta}^{-1} + \vSig_0^{-1}. \\
    \end{align*}
    
    \item For ${\ell} = 1, ..., L$ and $\vs = \vs_1, ..., \vs_n$, update $[\vbe_{\ell}(\vs_i)|\cdot] \sim Gau\left(\vV^{-1} \va, \vV^{-1}\right)$, where
    \begin{align*}
        \va & = \vX_{\ell}(\vs_i)'\vSig^{-1}_{\varepsilon_{\ell}}(\vs_i)\vz_{\ell}(\vs_i) + \vSig_{\beta}^{-1}\left(\vbe_{\ell-1}(\vs_i) + \vw_{\ell}(\vs_i)\right) + \vSig_{\beta}^{-1}\left(\vbe_{\ell+1}(\vs_i) + \vw_{\ell+1}(\vs_i)\right), \\
        \vV & = \vX_{\ell}(\vs_i)'\vSig^{-1}_{\varepsilon_{\ell}}(\vs_i)\vX_{\ell}(\vs_i) + 2\vSig_{\beta}^{-1}
    \end{align*}
    
    \item $[\vSig_{\beta}|\cdot] \sim IW(\Lambda, \Xi)$, where
    \begin{align*}
        \Lambda & = V + \sum_{i=1}^{n}\sum_{\ell=1}^L \left( \vbe_{\ell}(\vs_i) - \vbe_{\ell-1}(\vs_i) - \vw_{\ell}(\vs_i) \right)'\left( \vbe_{\ell}(\vs_i) - \vbe_{\ell-1}(\vs_i) - \vw_{\ell}(\vs_i) \right), \\
        \Xi & = nL + \xi.
    \end{align*}

    \item For $\vs = \vs_1, ..., \vs_n$ and $j = 1, 2$, update $[\sigma^2_{\varepsilon_j}(\vs_i)|\cdot] \sim IG(a_{\varepsilon}, b_{\varepsilon})$, where
    \begin{align*}
        a_{\varepsilon} = \frac{L*n}{2} + a \quad \text{and} \quad b_{\varepsilon} = b + \frac{1}{2}\sum_{\ell=1}^L(\vz_{j\ell}(\vs_i) - \vX_{\ell}\vbe_{j\ell}(\vs_i))'(\vz_{j\ell}(\vs_i) - \vX_{\ell}\vbe_{j\ell}(\vs_i)).
    \end{align*}
\end{enumerate}

\section*{Acknowledgements}
JSN and EMS were funded in part by the National Science Foundation \#EF-1638550. JSN was also partially funded by Missouri EPSCoR, supported by the National Science Foundation under Award \#IIA‑1355406 and \#IIA‑1430427. CKW was supported by NSF Award \#DMS-1811745.  The computation for this work was performed on the high performance computing infrastructure provided by Research Computing Support Services and in part by the National Science Foundation under grant number CNS-1429294 at the University of Missouri, Columbia Mo.  NCEP Reanalysis data provided by the NOAA/OAR/ESRL PSD, Boulder, Colorado, USA, from their Web site at https://www.esrl.noaa.gov/psd/.

\end{document}